\documentclass[graybox]{svmult}

\usepackage{type1cm}        % activate if the above 3 fonts are
\usepackage{makeidx}         % allows index generation
\usepackage{graphicx}        % standard LaTeX graphics tool
\usepackage{multicol}        % used for the two-column index
\usepackage[bottom]{footmisc}% places footnotes at page bottom

\usepackage{newtxtext}       % 
\usepackage{newtxmath}       % selects Times Roman as basic font
\usepackage{bm}

\makeindex             % used for the subject index
\begin{document}

\title*{An Interacting Agent Model of Economic Crisis}
\author{Yuichi Ikeda}
\institute{Yuichi Ikeda \at Graduate School of Advanced Integrated Studies in Human Survivability, 
Kyoto University, \email{ikeda.yuichi.2w@kyoto-u.ac.jp}}
\maketitle

\abstract{
Most national economies are linked by international trade. Consequently, economic globalization forms a massive and complex economic network with strong links, that is, interactions arising from increasing trade. 
Various interesting collective motions are expected to emerge from strong economic interactions in a global economy under trade liberalization. Among the various economic collective motions, economic crises are our most intriguing problem.
In our previous studies, we have revealed that the Kuramoto's coupled limit-cycle oscillator model and the Ising-like spin model on networks are invaluable tools for characterizing the economic crises.
In this study, we develop a mathematical theory to describe an interacting agent model that derives the coupled limit-cycle oscillator model and the Ising-like spin model by using appropriate approximations.
Our interacting agent model suggests phase synchronization and spin ordering during economic crises.
We confirm the emergence of the phase synchronization and spin ordering during economic crises by analyzing various economic time series data.
We also develop a network reconstruction model based on entropy maximization that considers the sparsity of the network. 
Here network reconstruction means estimating a network's adjacency matrix from a node's local information.
The interbank network is reconstructed using the developed model, and a comparison is made of the reconstructed network with the actual data. We successfully reproduce the interbank network and the known stylized facts.
In addition, the exogenous shock acting on an industry community in a supply chain network and financial sector are estimated. 
Estimation of exogenous shocks acting on communities of in the real economy in the supply chain network provide evidence of the channels of distress propagating from the financial sector to the real economy through the supply chain network. 
}

\section{Introduction}
\label{sec:1}

Most national economies are linked by international trade. Consequently, economic globalization forms a massive and complex economic network with strong links, that is, interactions due to increasing trade. In Japan, several small and medium enterprises would achieve higher economic growth by free trade based on the establishment of economic partnership agreement, such as the Trans-Pacific Partnership. 
Various collective motions exist in natural phenomena. For instance, a heavy nucleus that consists of a few hundred nucleons is largely deformed in a highly excited state and subsequently proceeds to nuclear fission. This phenomenon is a well-known example of quantum mechanical collective motion due to strong nuclear force between nucleons. 
From the analogy with the collective motions in natural phenomena, various interesting collective motions are expected to emerge because of strong economic interactions in a global economy under trade liberalization. Among the various economic collective motions, economic crises are our most intriguing problem.

\runinhead{Business Cycle} 

There have been several theoretical studies on the concept of the ``business cycle'' \cite{Haberler1937, Burns1964, Granger1964}. In recent times, the synchronization \cite{Huygens1966} of international business cycle as an example of the economic collective motion has attracted economists and physicists \cite{Krugman1966}. Synchronization of business cycles across countries has been discussed using correlation coefficients between GDP time series \cite{Stock2005}. However, this method remains only a primitive first step, and a more definitive analysis using a suitable quantity describing business cycles is needed.
In an analysis of business cycles, an important question is the significance of individual (micro) versus aggregate (macro) shocks. Foerster et al. \cite{Foerster2011} used factor analysis to show that the volatility of U.S. industrial production was largely explained by aggregate shocks and partly by cross-sectoral correlations from the individual shocks transformed through the trade linkage. 
The interdependent relationship of the global economy has become stronger because of the increase in international trade and investments \cite{Tzekinaa2008, Barigozzi2011, He2010, Piccardi2012}. 

We took a different approach to analyze the shocks to explain the synchronization in international business cycles.
We analyzed the quarterly GDP time series for Australia, Canada, France, Italy, the United Kingdom, and the United States from Q2 1960 to Q1 2010 to determine the synchronization in international business cycles \cite{Ikeda2013a}. The followings results were obtained.

(1) The angular frequencies $\omega_i$ estimated using the Hilbert transform are almost identical for the six countries. Therefore, frequency entrainment is observed. Moreover, the phase locking indicator $\sigma(t)$ shows that partial phase locking is observed for the analyzed countries, representing direct evidence of synchronization in international business cycles.

(2) A coupled limit-cycle oscillator model was developed to explain the synchronization mechanism. A regression analysis showed that the model is a very good fit for the phase time series of the GDP growth rate. The validity of the model implies that the origin of the synchronization is the interaction resulting from international trade.

(3) We also showed that information from economic shocks is carried by phase time series $\theta_i(t)$. The comovement and individual shocks are separated using the random matrix theory. A natural interpretation of the individual shocks is that they are ``technological shocks''. The present analysis demonstrates that average phase fluctuations well explain business cycles, particularly recessions. Because it is highly unlikely that all of the countries are subject to common negative technological shocks, the results obtained suggest that pure ``technological shocks'' cannot explain business cycles.

(4) Finally, the obtained results suggest that business cycles may be understood as comovement dynamics described by the coupled limit-cycle oscillators exposed to random individual shocks. The interaction strength in the model became large in parallel with the increase in the amounts of exports and imports relative to GDP. Therefore, a significant part of comovements comes from international trade.

We observed various tpes of collective motions for economic dynamics, such as synchronization of business cycles \cite{Ikeda2013a, Ikeda2013b, Ikeda2014}, on the massive complex economic network.
The linkages among national economies play important roles in economic crises and during normal economic states. 
Once an economic crisis occurs in a certain country, the influence propagates instantaneously toward the rest of the world.
For instance, the global economic crisis initiated by the bankruptcy of Lehman Brothers in 2008 is still fresh in our minds.
The massive and complex global economic network might show characteristic collective motion during economic crises.

\runinhead{Economic Crisis} 

Numerous preceding studies attempted to explain the characteristics of stock market crash using spin variables in the econophysics literature. First, we note some content and mathematical descriptions from previous studies \cite{Sornette2014}, \cite{Bouchaud2013}. In particular, we note studies by Kaizoji and Sornette \cite{Kaizoji2002}, \cite{Kaizoji2000}, \cite{Bornholdt2001}, \cite{Sornette2006}, \cite{Harras2011}, \cite{Vikram2011}, \cite{Johansen2000}, \cite{Nadal2005} in which investor strategies (buy or sell) are modeled as spin variables, with stock prices varying depending on the differences in the number of spin-ups. In addition, the feedback effect on an investor's decision making through a neighbor's strategies can explain bubble formations and crashes. For instance, the temporal evolution is simulated by adding random components in Sornette and Zhou \cite{Sornette2006}. Most papers adopted two-state spin variables; however, the study by Vikram and Sinha \cite{Vikram2011} adopted three-state spin variables. Note that the purpose of these studies was to reproduce the scaling law and not explain phase transitions.

In contrast, economics journals aim to explain the optimality of investors' decision making \cite{Harras2011}. In Nadal et al. \cite{Nadal2005}, phase transition is discussed, starting with discrete choice theory. Many papers have similar discussions on phase transitions, with slight variations in optimization and profit maximization. Although empirical studies using real data are relatively few, Wall Street market crash in 1929, 1962, and 1987 and the Hong Kong Stock Exchange crash in 1997 were studied in Johansen, Ledoit, and Sornette \cite{Johansen2000}. Note that elaborate theoretical studies exist on phase transition effects on networks and the thermodynamics of networks \cite{Aleksiejuk2002}, \cite{Dorogovtsev2002}, \cite{Dorogovtsev2008}, \cite{Ye2015}.
Furthermore, preceding studies on macroprudential policy exist that mainly focus on time series analyses of macroeconomic variables \cite{Borio2012}, \cite{Borio2014}, \cite{Borio2011}.

Although the market crash is an important part of an economic crisis, our main interest is that the real economy consists of many industries in various countries.
We analyzed industry-sector-specific international trade data to clarify the structure and dynamics of communities that consist of industry sectors in various countries linked by international trade \cite{Ikeda2016}. We applied conventional community analysis to each time slice of the international trade network data: the World Input Output Database. This database contains industry-sector-specific international trade data on 41 countries and 35 industry sectors from 1995 to 2011. Once the community structure was obtained for each year, the links between communities in adjoining years were identified using the Jaccard index as a similarity measure between communities in adjoining years.

The identified linked communities show that six backbone structures exist in the international trade network. The largest linked community is the Financial Intermediation sector and the Renting of Machines and Equipments sector in the United States and the United Kingdom. The second is the Mining and Quarrying sector in the rest of the world, Russia, Canada, and Australia. The third is the Basic Metals and Fabricated Metal sector in the rest of the world, Germany, Japan, and the United States. These community structures indicate that international trade is actively transacted among the same or similar industry sectors. Furthermore, the robustness of the observed community structure was confirmed by quantifying the variations in the information for perturbed network structure.
The theoretical study conducted using a coupled limit-cycle oscillator model suggests that the interaction terms from international trade can be viewed as the origin of the synchronization. 

The economic crisis of 2008 showed that the conventional microprudential policy to ensure the soundness of individual banks was not sufficient, and prudential regulations to cover the entire financial sector were desired. Such regulations attract increasing attention, and policies related to such regulations are called a macroprudential policy that aims to reduce systemic risk in the entire financial sector by regulating the relationship between the financial sector and the real economy. 
We studied channels of distress propagation from the financial sector to the real economy through the supply chain network in Japan from 1980 to 2015 using a Ising-like spin model on networks  \cite{Ikeda2018}.
An estimation of exogenous shocks acting on communities of the real economy in the supply chain network provided evidence of channels of distress propagation from the financial sector to the real economy through the supply chain network. Causal networks between exogenous shocks and macroeconomic variables clarified the characteristics of the lead lag relationship between exogenous shocks and macroeconomic variables when the bubble bursts.

\section{Interacting Agent Models}
\label{sec:2}

The coupled limit-cycle oscillator model and the Ising-like spin model on networks are invaluable tools to characterize an economic crisis, as described in Section \ref{sec:1}.
In this section, we develop a mathematical theory to describe an interacting agent model that derives the aforementioned two models using appropriate approximations.

\subsection{Interacting Agent Model on Complex Network}
\label{subsec:2.1}

\runinhead{Hamiltonian Dynamics} 

Our system consists of $N$ company agents and $M$ bank agents. The states of the agents are specified by multi-dimensional state vectors $\bm{q}_i$ and $\bm{q}_j$ for companies and banks, respectively. 
If we consider (1) security indicators: (1-1) total common equity divided by total assets and (1-2) fixed assets divided by total common equity; (2) profitability indicators: (2-1) operating income divided by total assets and (2-2) operating income divided by total revenue; (3) capital efficiency indicator: total revenue divided by total assets, and (4) growth indicator: operating income at time $t$ divided by operating income at time $t-1$ as variables to the soundness of companies,  state vectors $\bm{q}_i$ are expressed in six dimensional space.
The agents interact in the following way:
\begin{equation}
  H_{int} (\bm{q}) = - \sum_{i \in C} \bm{H}_{C,i} \bm{q}_i - \sum_{j \in B} \bm{H}_{B,j} \bm{q}_j -  J_C \sum_{i \in C,j \in C} a_{ij} \bm{q}_i \bm{q}_j - J_{CB} \sum_{i \in C,j \in B}  b_{ij} \bm{q}_i \bm{q}_j,
\label{eq:Hamiltonian_q}
\end{equation}
where $H_C$, $H_B$, $J_C$, $J_{CB}$, $a_{ij}$, and $b_{ij}$ represent the exogenous shock acting on companies, the exogenous shock acting on banks, the strength of the inter-company interactions, the strength of company-bank interactions, the adjacent matrix of the supply chain network, and the adjacent matrix of the bank to company lending network, respectively.
The Hamiltonian $H(\bm{q} ,\bm{p})$ of the system is the sum of the kinetic energy of th e companies (the first term), the kinetic energy of the banks (the second term), and the interaction potential $H_{int} (\bm{q})$:  
\begin{equation}
  H(\bm{q} ,\bm{p}) = \sum_{i \in C} \frac{\bm{p}_i^2}{2m} + \sum_{j \in B} \frac{\bm{p}_j^2}{2m} + H_{int} (\bm{q}),
\label{eq:Hamiltonian}
\end{equation}
where $\bm{p}_i$ and $m$ are the multi-dimensional momentum vector and the mass of agent $i$. Here, we set $m=1$ without loss of generality. 
We obtain the canonical equations of motion for agent $i$:
\begin{equation}
  \frac{\partial H(\bm{q}, \bm{p})}{\partial \bm{p}_i} = \bm{\dot q}_i,
\label{eq:EOM1}
\end{equation}
\begin{equation}
  \frac{\partial H(\bm{q}, \bm{p})}{\partial \bm{q}_i} = - \bm{\dot p}_i.
\label{eq:EOM2}
\end{equation}
From Eq. (\ref{eq:EOM1}), we obtain
\begin{equation}
  \frac{\bm{p}_i}{m} = \bm{\dot q}_i.
\label{eq:EOM1p}
\end{equation}
By substituting Eq. (\ref{eq:EOM1p}) into Eq. (\ref{eq:EOM2}), we obtain the equation of motion of the company agent $i$: 
\begin{equation}
   \bm{\ddot q}_i = \bm{H}_{C,i} + J_C \sum_{j \in C} \left( a_{ij} + a_{ji} \right) \bm{q}_j + J_{CB} \sum_{j \in B} b_{ij} \bm{q}_j,
\label{eq:EOM3}
\end{equation}
and the equation of motion of bank agent $j$:
\begin{equation}
   \bm{\ddot q}_j = \bm{H}_{B,j} + J_{CB} \sum_{i \in C} b_{ij} \bm{q}_i.
\label{eq:EOM4}
\end{equation}

\runinhead{Langevin Dynamics} 

Hamiltonian dynamics are applicable to a system in which its total energy is conserved. 
However, to be noted is that the economic system is open and no quantity is conserved accurately.
Considering this point, we add a term for constant energy inflow $\bm{P}_i$ and a term for energy dissipation outside the system $- \alpha \bm{\dot q}_i$ to the equation of motion for company agent $i$ in Eq. (\ref{eq:EOM3}): 
\begin{equation}
   \bm{\ddot q}_i = \bm{P}_i - \alpha \bm{\dot q}_i + \bm{H}_{C,i} + J_C \sum_{j \in C} \left( a_{ij} + a_{ji} \right) \bm{q}_j + J_{CB} \sum_{j \in B} b_{ij} \bm{q}_j,
\label{eq:LD1}
\end{equation}
where we assume that energy dissipation is proportional to the velocity of agent $\bm{\dot q}_i$.
The stochastic differential equation in Eq. (\ref{eq:LD1}) is called the Langevin equation.

Similarly, way, we obtain the Langevin equation for bank agent $j$:
\begin{equation}
   \bm{\ddot q}_j = \bm{P}'_j - \alpha' \bm{\dot q}_j + \bm{H}_{B,j} + J_{CB} \sum_{i \in C} b_{ij} \bm{q}_i,
\label{eq:LD2}
\end{equation}

If a system's inertia is very large and thus $\bm{\ddot q}_i \simeq 0$, we obtain the following first order stochastic differential equation for company agent $i$:
\begin{equation}
   \bm{\dot q}_i = \frac{\bm{P}_i}{\alpha} + \frac{\bm{H}_{C,i}}{\alpha} + \frac{J_C}{\alpha} \sum_{j \in C} \left( a_{ij} + a_{ji} \right) \bm{q}_j + \frac{J_{CB}}{\alpha} \sum_{j \in B} b_{ij} \bm{q}_j.
\label{eq:LD3}
\end{equation}
Similarly, we obtain the  first order stochastic differential equation for bank agent $j$:
\begin{equation}
   \bm{\dot q}_j = \frac{\bm{P}'_j}{\alpha'} + \frac{\bm{H}_{B,j}}{\alpha'} + \frac{J_{CB}}{\alpha'} \sum_{i \in C} b_{ij} \bm{q}_i.
\label{eq:LD4}
\end{equation}

\subsection{Ising Model with Exogenous Shock}
\label{subsec:2.2}

\runinhead{Underlying Approximated Picture} 

Suppose that $\bm{q}_i$ is a one dimensional variable, and we assume that magnitude $|q_i|$ varies slowly compared with orientation $s_i$. 
We approximate the magnitude as $|q_i| \approx const.$ and obtain:

\begin{equation}
   s_i = \mathrm{sgn} \frac{q_i}{|q_i|}, 
\label{eq:spin}
\end{equation}
\begin{equation}
   \frac{\partial}{\partial q_i} = \frac{\partial}{\partial s_i} \frac{\partial s_i}{\partial q_i} = \frac{1}{|q_i|} \frac{\partial}{\partial s_i}.
\label{eq:del_spin}
\end{equation}

\runinhead{Derived Model} 

Stock price $x_{i,t} (i=1,\cdots,N (or M), t=1,\cdots,T)$ is assumed to be a surrogate variable to indicate the soundness of companies or banks. The one-dimensional spin variable $s_{i,t}$ was estimated from the log return of daily stock prices, $r_{i,t}$:
\begin{equation}
s_{i,t}=+1	\quad \left( q_{i,t} = \log x_{i,t} - \log x_{i,t-1} \ge0 \right)
\label{eq:SpinUp}
\end{equation}
\begin{equation}
s_{i,t}=-1	\quad \left( q_{i,t} = \log x_{i,t} - \log x_{i,t-1} <0 \right)
\label{eq:SpinDown}
\end{equation}
Here spin-up: $s_{i,t}=+1$ indicates that company $i$ is in good condition; on spin-down: $s_{i,t}=-1$ indicates that company $i$ is in bad condition. The macroscopic order parameter $M_t = \sum_i s_{i,t}$  is an indicator of the soundness of the macro economy, which is regarded as an extreme simplification to capture the soundness of the economy. In addition to this simplification, spin variables include noise information because we have various distortions in the stock market caused by irrational investor decision making.
The spin variables of companies interact with the spins of other companies through the supply chain network and interact with banks through the lending network. Those interactions between companies and banks are mathematically expressed as Hamiltonian. A Hamiltonian is written as follows:
\begin{equation}
 H_{int} (s) = -H_C \sum_{i \in C} s_{i,t} - H_B \sum_{i \in B} s_{i,t} - J_C \sum_{i \in C,j \in C} a_{ij} s_{i,t} s_{j,t} - J_{CB} \sum_{i \in C,j \in B} a_{ij} s_{i,t} s_{j,t}
\label{eq:Hamiltonian1}
\end{equation}
where $H_C$ and $H_B$ are the exogenous shocks acting on companies and banks, respectively. $a_{ij}$ represents the elements of adjacency matrix $A$ of the supply chain network that is treated as a binary directed network. 

When spins are exposed for exogenous shock $H_{ext}$, an effective shock of
\begin{equation}
 H_{eff} = H_{ext} + H_{int}
\label{eq:EffShock}
\end{equation}
acts on each spin. By calculating interaction $H_{int}$ of the Hamiltonian of Eq. (\ref{eq:Hamiltonian1}), exogenous shock $H_{ext}$ was estimated by considering the nearest neighbor companies in the supply chain network
\begin{equation}
 \frac{M_t}{N\mu}  = \tanh⁡ \left( \frac{\mu H_{ext}}{kT} + \frac{J}{kT} \frac{\sum_{ij} (a_{ij} + a_{ji}) s_{i,t}}{N} \mu \right)
\label{eq:shock2}
\end{equation}
where $T$ represents temperature as a measure of the activeness of the economy, which is considered proportional to GDP per capita.
We note that supply chain network data are a prerequisite condition for estimating exogenous shock $H_{ext}$.

In the current model, the interactions between banks
\begin{equation}
  \sum_{i \in B,j \in B} J_{BB} t_{ij} s_{i,t} s_{j,t}
\label{eq:Bank}
\end{equation}
were ignored because of a lack of data for interbank network $t_{ij}$. This lack of data caused by the central bank not making public the data on transactions between banks. 
A method to reconstruct the interbank network is described in Section \ref{sec:3}.

\subsection{Kuramoto Model with Exogenous Shock} %
\label{subsec:2.3}

\runinhead{Underlying  Approximated Picture} 

When $\bm{q}_i$ is a two-dimensional variable, we treat this quantity as if it is a complex variable. We assume that amplitude $|\bm{q}_i|$ varies slowly compared with phase $\theta_i$. We approximate the amplitude as $|\bm{q}_i| \approx const.$ and, thus, obtain:

\begin{equation}
   \bm{q}_i = |\bm{q}_i| e^{i \theta_i},
\label{eq:pahse}
\end{equation}
\begin{equation}
   \frac{\partial}{\partial \bm{q}_i} = \frac{\partial}{\partial \theta_i} \frac{\partial \theta_i}{\partial \bm{q}_i}.
\label{eq:del_theta}
\end{equation}

\runinhead{Derived Model} 

The business cycle is observed in most industrialized economies. Economists have studied this phenomenon by means of mathematical models, including various types of linear, nonlinear, and coupled oscillator models.

Interdependence, or coupling, between industries in the business cycle has been studied for more than half a century.
A study of the linkages between markets and industries using nonlinear difference equations suggests a dynamical coupling among industries \cite{Goodwin1947}.
A nonlinear oscillator model of the business cycle was then developed using a nonlinear accelerator as the generation mechanism \cite{Goodwin1951}.
We stress the necessity of nonlinearity because linear models are unable to reproduce sustained cyclical behavior, and tend to either die out or diverge to infinity. 

However, a simple linear economic model, based on ordinary economic principles, optimization behavior, and rational expectations, can produce cyclical behavior much like that found in business cycles \cite{LongPlosser1983}. 
An important question aside from synchronization in the business cycle is whether sectoral or aggregate shocks are responsible for the observed cycle. This question was empirically examined; it was found that business cycle fluctuations are caused by small sectoral shocks, rather than by large common shocks \cite{LongPlosser1987}.

As the third model category, coupled oscillators were developed to study noisy oscillating processes such as national economies \cite{Anderson1999} \cite{Selover2003}.
Simulations and empirical analyses showed that synchronization between the business cycles of different countries is consistent with such mode-locking behavior. 
Along this line of approach, a nonlinear mode-locking mechanism was further studied that described a synchronized business cycle between different industrial sectors \cite{Sussmuth2003}.

Many collective synchronization phenomena are known in physical and biological systems \cite{Strogatz2000}.
Physical examples include clocks hanging on a wall, an array of lasers, microwave oscillators, and Josephson junctions.
Biological examples include synchronously flashing fireflies, networks of pacemaker cells in the heart, and metabolic synchrony in yeast cell suspensions.

Kuramoto established the coupled limit-cycle oscillator model to explain this wide variety of synchronization phenomena \cite{Kuramoto1975} \cite{Strogatz2000} \cite{Acebron2005}. 
In the Kuramoto model, the dynamics of the oscillators are governed by
\begin{equation}
\dot{\theta_i} = \omega_i + \sum_{j=1}^{N}  k_{ji} \sin(\theta_j - \theta_i),
\label{eq:Kuramoto}
\end{equation}
where $\theta_i$, $\omega_i$, and $k_{ji}$ are the oscillator phase, the natural frequency, and the coupling strength, respectively.
If the coupling strength $k_{ij}$ exceeds a certain threshold that equals the natural frequency $\omega_i$, the system exhibits synchronization.

By explicitly writing amplitude $|q_i|$ and phase $\theta_i$, the third term of the R.H.S. in Eq. (\ref{eq:Hamiltonian_q}) is rewritten as follows:
\begin{equation}
 J_C \sum_{i \in C,j \in C} a_{ij} \bm{q}_i \bm{q}_j = J_C \sum_{i \in C,j \in C} a_{ij} |\bm{q}_j| |\bm{q}_i| \cos(\theta_j - \theta_i). 
\label{qiqj}
\end{equation}
The spatial derivative of Eq. (\ref{qiqj}) is obtained:
\begin{equation}
\begin{split}
 \frac{\partial}{\partial \bm{q}_i } J_C \sum_{i \in C,j \in C} a_{ij} \bm{q}_i \bm{q}_j &= \frac{\partial \theta_i}{\partial \bm{q}_i} \frac{\partial}{\partial \theta_i} J_C \sum_{i \in C,j \in C} a_{ij} |\bm{q}_j| |\bm{q}_i| \cos(\theta_j - \theta_i) \\
 &= \frac{\partial \theta_i}{\partial \bm{q}_i} J_C \sum_{j \in C} \left( a_{ij} + a_{ji} \right) |\bm{q}_j| |\bm{q}_i| \sin(\theta_j - \theta_i).
\end{split}
\label{qiqj_force}
\end{equation}
By substituting Eq. (\ref{qiqj_force}) into the stochastic differential equation for company agent $i$ of Eq. (\ref{eq:LD3}), we obtain the following equation:
\begin{equation}
\begin{split}
   \frac{\partial \bm{q}_i}{\partial \theta_i} \frac{d \theta_i}{dt}  &= \frac{\bm{P}_i}{\alpha} + \frac{\bm{H}_{C,i}}{\alpha}  \\
   &+ \frac{J_C}{\alpha} \frac{\partial \theta_i}{\partial \bm{q}_i} \sum_{j \in C} \left( a_{ij} + a_{ji} \right) |\bm{q}_j| |\bm{q}_i| \sin(\theta_j - \theta_i) \\
   &+ \frac{J_{CB}}{\alpha} \frac{\partial \theta_i}{\partial \bm{q}_i} \sum_{j \in B} b_{ij} |\bm{q}_j| |\bm{q}_i| \sin(\theta_j - \theta_i).
\end{split}
\label{eq:LD3_phase}
\end{equation}
Consequently, we obtain the stochastic differential equation, which is equivalent to the Kuramoto model of Eq. (\ref{eq:Kuramoto}) with an additional exogenous shock term:
\begin{equation}
\begin{split}
     \frac{d \theta_i}{dt}  &= \frac{1}{|\bm{q}_i|^2} \left( \frac{\bm{P}_i}{\alpha} + \frac{\bm{H}_{C,i}}{\alpha} \right) \frac{\partial \bm{q}_i}{\partial \theta_i} \\
   &+ \frac{J_C}{\alpha} \sum_{j \in C} \left( a_{ij} + a_{ji} \right) \frac{|\bm{q}_j|}{|\bm{q}_i|} \sin(\theta_j - \theta_i) \\
   &+ \frac{J_{CB}}{\alpha} \sum_{j \in B} b_{ij} \frac{|\bm{q}_j|}{|\bm{q}_i|} \sin(\theta_j - \theta_i).
\end{split}
\label{eq:LD3_phase2}
\end{equation}

\section{Network Reconstruction}
\label{sec:3}

Network reconstruction estimates a network's adjacency matrix from a node's local information.
We developed a network reconstruction model based on entropy maximization and considering network sparsity.

\subsection{Existing Models}
\label{sec:3.1}

\runinhead{MaxEnt algorithm}

The MaxEnt algorithm maximizes entropy $S$ by changing $t_{ij}$ under the given total lending $s_{i}^{out}$ and the total borrowing $s_{i}^{in}$ for bank $i$ \cite{Wells2004}, \cite{Upper2011}.
The analytical solution of this algorithm is easily obtained as
\begin{equation}
 t_{ij}^{ME} = \frac{s_{i}^{out} s_{j}^{in}}{G},
\label{MaxEnt4}
\end{equation}
\begin{equation}
 G = \sum_i s_{i}^{out} = \sum_j s_{j}^{in}.  
\label{MaxEnt5}
\end{equation}
However, noted is that the solution to Eq. (\ref{MaxEnt4}) provides a fully connected network, although real-world networks are often known as sparse networks.

\runinhead{Iterative proportional fitting} 

Iterative proportional fitting (IPF) has been introduced to correct the dense property of $t_{ij}^{ME}$ at least partially.
By minimizing the Kullback-Leibler divergence between a generic nonnegative $t_{ij}$ with null diagonal entries and the MaxEnt solution $t_{ij}^{ME}$ in Eq. (\ref{MaxEnt4}),
we obtain $t_{ij}^{IPF}$ \cite{Squartini2018}:
\begin{equation}
 \min \left( \sum_{ij (i \neq j)} t_{ij} \ln \frac{t_{ij}}{t_{ij}^{ME}} \right) = \sum_{ij (i \neq j)} t_{ij}^{IPF} \ln \frac{t_{ij}^{IPF}}{t_{ij}^{ME}}.  
\label{IPF}
\end{equation}
The solution $t_{ij}^{IPF}$ has null diagonal elements, but does show the sparsity equivalent to real-world networks.

\runinhead{Drehmann and Tarashev approach}

Starting from the MaxEnt matrix $t_{ij}^{ME}$, a sparse network is obtained in the following three steps \cite{Drehmann2013}: 
First, choose a random set of off-diagonal elements to be zero.
Second, treat the remaining nonzero elements as random variables distributed uniformly between zero and twice their MaxEnt estimated value $t_{ij}^{DT} \sim U(0, 2t_{ij}^{ME})$.
Therefore, the expected value of weights under this distribution coincides with the MaxEnt matrix $t_{ij}^{ME}$.
Third, the IPF algorithm is run to correctly restore the value of the total lending $s_{i}^{out}$ and the total borrowing $s_{i}^{in}$. 
However, note that we need to specify the set of off-diagonal nonzero elements. 
Therefore, accurate sparsity does not emerge spontaneously in this approach.

\subsection{Ridge Entropy Maximization Model}
\label{sec:3.2}

\runinhead{Convex Optimization} 
\label{Convex}

We develop a reconstruction model for the economic network and apply it to the interbank network in which nodes and links are banks and lending or borrowing amounts, respectively. 
First, we maximize configuration entropy $S$ under the given total lending $s_{i}^{out}$ and total borrowing $s_{i}^{in}$ for bank $i$.

Configuration entropy $S$ is written using bilateral transaction $t_{ij}$ between bank $i$ and $j$ as follows,
\begin{equation}
 S = \log \frac{ \left( \sum_{ij} t_{ij} \right) ! }{ \prod_{ij} t_{ij} ! } \approx \left( \sum_{ij} t_{ij} \right) \log \left( \sum_{ij} t_{ij} \right) - \sum_{ij} t_{ij} \log t_{ij}.
\label{entropy1}
\end{equation}
Here, an approximation is applied to the factorial $!$ using Stirling's formula.   
The first term of the R.H.S. of Eq.(\ref{entropy1}) does not change the value of $S$ by changing $t_{ij}$ because $\sum_{ij} t_{ij}$ is constant. Consequently, we have a convex objective function:
\begin{equation}
 S = - \sum_{ij} t_{ij} \log t_{ij}.
\label{entropy2}
\end{equation}
Entropy $S$ is to be maximized with the following constraints:
\begin{equation}
 s_{i}^{out} = \sum_j t_{ij},
\label{TotExp}
\end{equation}
\begin{equation}
 s_{j}^{in} = \sum_i t_{ij},
\label{TotImp}
\end{equation}
\begin{equation}
 G = \sum_{ij} t_{ij}.
\label{TotTrade}
\end{equation}
Here, constraints Eq. (\ref{TotExp}) and Eq. (\ref{TotImp}) correspond to local information about each node.

\runinhead{Sparse Modeling} 
\label{Sparse}

The accuracy of the reconstruction will be improved using the sparsity of the interbank network.
We have two different types of sparsity here. 
The first is characterized by the skewness of the observed bilateral transaction distributions. The second type of sparsity is characterized by the skewness of the observed in-degree and out-degree distributions. Therefore, a limited fraction of nodes have a large number of links, and most nodes have a small number of links. Consequently, the adjacency matrix of international trade is sparse.

To take into account the first type of sparsity,
the objective function of Eq. (\ref{entropy2}) is modified by applying the concept of Lasso (least absolute shrinkage and selection operator) \cite{Tibshirani1996}  \cite{Breiman1995} \cite{Hastie2008} to our convex optimization problem. 

By considering this fact, our problem is reformulated as the maximization of objective function $z$:
\begin{equation}
 z(t_{ij}) = S - \sum_{ij} t_{ij}^2 = - \sum_{ij} t_{ij} \log t_{ij} - \beta \sum_{ij} t_{ij}^2
\label{Ridge}
\end{equation}
with local constraints. Here the second term of R.H.S. of Eq. (\ref{Ridge}) is L2 regularization.

\runinhead{Ridge Entropy Maximization Model} 
\label{Thermodynamics}

In the theory of thermodynamics, a system's equilibrium is obtained by minimizing thermodynamic potential $F$:
\begin{equation}
 F = E - TS
\label{FreeEnergy}
\end{equation}
where $E$, $T$, and $S$ are internal energy, temperature, and entropy, respectively. Eq. (\ref{FreeEnergy}) is rewritten as a maximization problem as follows:
\begin{equation}
 z \equiv -\frac{1}{T} F = S - \frac{1}{T} E.
\label{ObjtFcun}
\end{equation}
We note that Eq. (\ref{ObjtFcun}) has the same structure as Eq. (\ref{Ridge}). Thus, we interpret the meaning of control parameter $\beta$ and L2 regularization as inverse temperature and internal energy, respectively.
In summary, we have a ridge entropy maximization model \cite{Ikeda2018b} \cite{Ikeda2020}: 
\begin{equation}
\renewcommand{\arraystretch}{2.3}
%\begin{displaymath}
\begin{array}{ll}
\mbox{maximize   } & z (p_{ij}) = - \sum_{ij} p_{ij} \log p_{ij} - \beta \sum_{ij} p_{ij}^2 \\ 
\mbox{subject to   } & G = \sum_{ij} t_{ij} \\
                       & \frac{s_{i}^{out}}{G} = \sum_j \frac{t_{ij}}{G} = \sum_j p_{ij} \\ 
                       & \frac{s_{j}^{in}}{G} = \sum_i \frac{t_{ij}}{G} = \sum_i p_{ij} \\
                       & t_{ij} \geq 0 \\ 
\end{array}
%\end{displaymath}
\label{ConvexFormulation}
\end{equation}

\section{Empirical Validation of Models}
\label{sec:4}

The model described in Section $\ref{sec:2}$ suggests the phase synchronization and the spin ordering during an economic crisis.
In this section, we confirm the phase synchronization and the spin ordering by analyizing varilous economic time series data.
In addition, the exogenous shock acting on an industry community in a supply chain network and the financial sector are estimated. 
An estimation of exogenous shocks acting on communities of the real economy in the supply chain network provided evidence of the channels of distress propagation from the financial sector to the real economy through the supply chain network. 
Finally, we point out that the interactions between banks were ignored in the interacting agent model explained in Section $\ref{subsec:2.2}$ given a lack of transaction data $t_{ij}$ in an interbank network. 
This lack of data is caused by the central bank not making public the data on transactions between banks. 
In this section, the interbank network is reconstructed and the reconstructed network is compared with the actual data and the known stylized facts.

\subsection{Phase Synchronization and Spin Ordering during Economic Crises}

\begin{figure*}
  \includegraphics[width=0.8\textwidth]{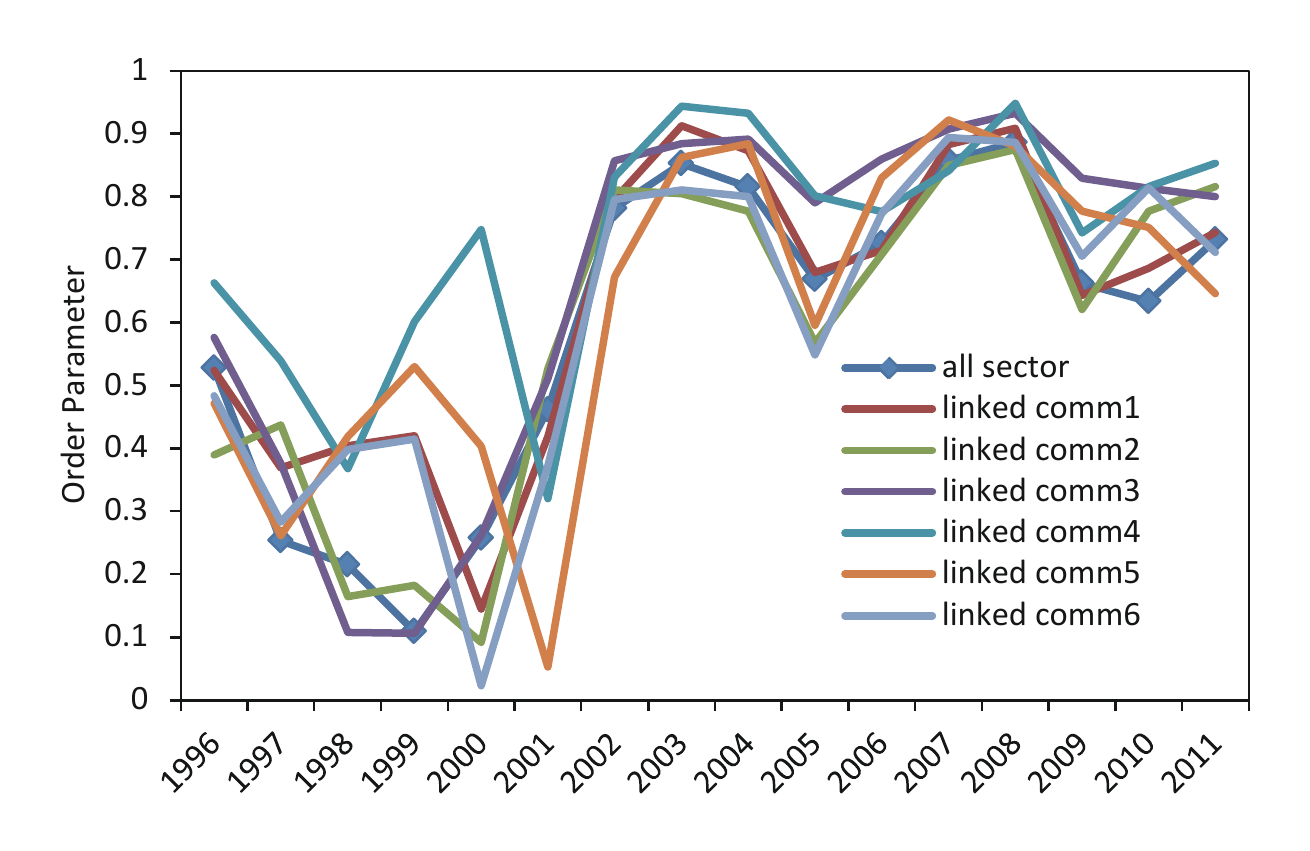}
\caption{Temporal change in amplitude for the order parameters $r(t)$ of phase synchonization for each community: We applied a conventional community analysis to each time slice of the international trade network data: the World Input Output Database. This database contains the industry-sector-specific international trade data on 41 countries and 35 industry sectors from 1995 to 2011. Once the community structure was obtained for each year, the links between communities in adjoining years was identified by using the Jaccard index as a similarity measure between communities in adjoining years.}
\label{fig:OrderParameter}       % Give a unique label
\end{figure*}
\begin{figure*}
  \includegraphics[width=0.8\textwidth]{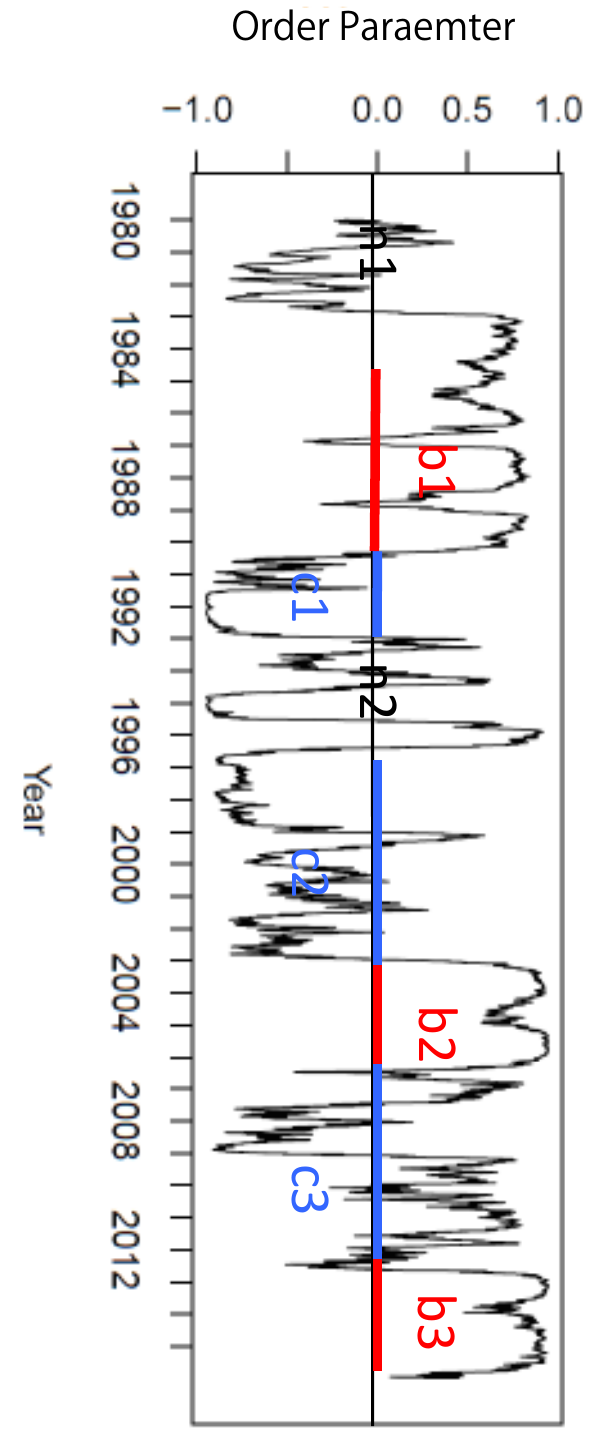}
\caption{Temporal changes of the order parameter of spin ordering for the real economy (companies) $M_C(t)$: The order parameter shows the high spin ordering $M_{C,t} \approx 1$ during the bubble periods, and $M_{C,t} \approx -1$ during the crisis periods. 
}
\label{fig:mag_company}       % Give a unique label
\end{figure*}
\begin{figure*}
  \includegraphics[width=0.8\textwidth]{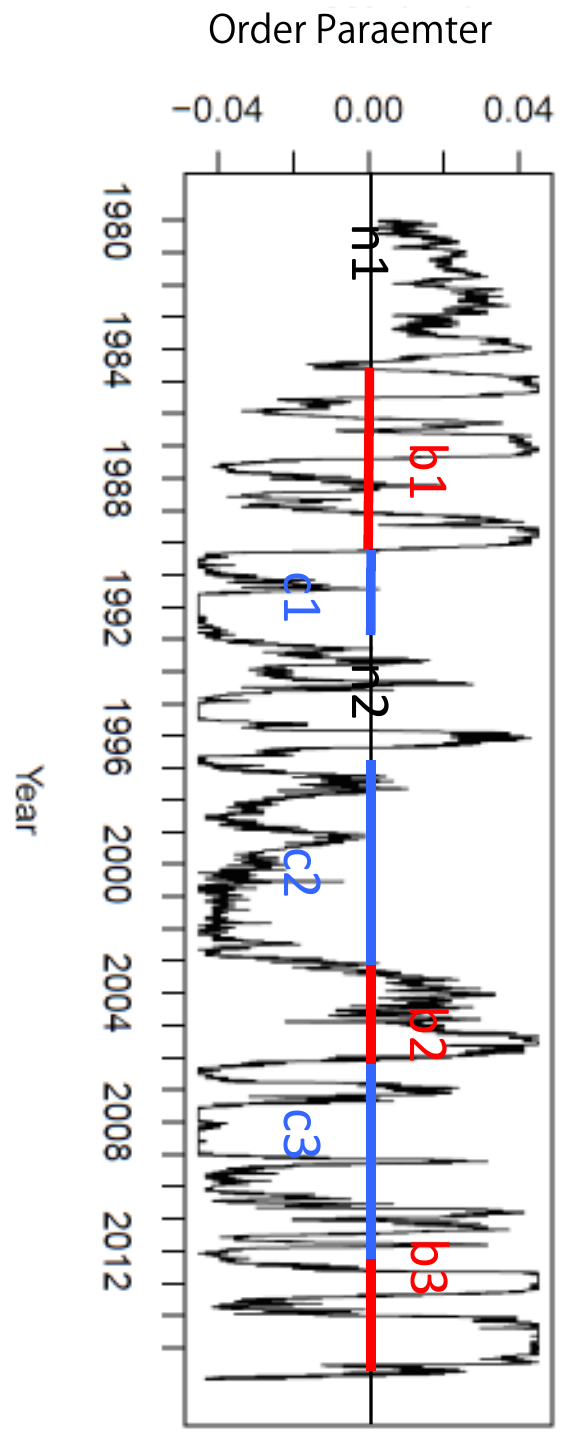}
\caption{Temporal changes of the order parameter of spin ordering for financial sector (banks) $M_B(t)$: The order parameter shows the high spin ordering $M_{B,t} \approx 1$ during the bubble periods, and $M_{B,t} \approx -1$ during the crisis periods. 
}
\label{fig:mag_bank}       % Give a unique label
\end{figure*}

We evaluated the phase time series of the growth rate of value added for 1435 nodes (41 countries and 35 industry sectors) from 1995 to 2011in the World Input Output Database) using the Hilbert transform and then estimated the order parameters of the phase synchronization for communities \cite{Ikeda2016}. 
The order parameter of the phase synchronization is defined by
\begin{equation}
 u(t) = r(t) e^{i \phi(t)} = \frac{1}{N} \sum_{j=1}^N e^{i \theta_{j}(t)}.
\label{PhaseOrderParameter}
\end{equation}
The respective amplitude for the order parameter of each community was observed to be greater than the amplitude for all sectors. 
Therefore, active trade produces higher phase coherence within each community. The temporal change in amplitude for the order parameters for each community are shown in Fig. \ref{fig:OrderParameter} from 1996 to 2011. 
Phase coherence decreased gradually in the late 1990s but increased sharply in 2002. From 2002, the amplitudes for the order parameter remained relatively high.
In particular, from 2002 to 2004, and from 2006 to 2008, we observe high phase coherence.
The first period was right after the dot-com crash and lasted from March 2000 to October 2002.
The second period corresponds to the subprime mortgage crisis that occurred between December 2007 and early 2009.
These results are consistent with the results obtained in the previous study \cite{Ikeda2013a}.

The stock price is the daily time series for the period from January 1, 1980 to December 31, 2015. The spin variable $s_{i,t}$ was estimated using Eqs. (\ref{eq:SpinUp}) and (\ref{eq:SpinDown}). 
The order parameters of spin ordering are defined by:
\begin{equation}
 M_{C,t} = \sum_{i \in C} s_{i,t}.
\label{SpinOrderParameterCompany}
\end{equation}
\begin{equation}
 M_{B,t} = \sum_{i \in B} s_{i,t}.
\label{SpinOrderParameterBank}
\end{equation}
for the real economy and the financial sector, respectively.
Temporal changes in the order parameter of spin ordering for the real economy (companies) and the financial sector (banks) are shown in Figs. \ref{fig:mag_company}  and \ref{fig:mag_bank}, respectively. 
The symbols in Figs. n1, b1, c1, n2, c2, b2, c3, and b3 show ``Normal period: 1980~-~1985'', ``Bubble period: 1985~-~1989'', ``Asset bubble crisis: 1989~-~1993'', ``Normal period: 1993~-~1997'', ``Financial crisis: 1997~-~2003'', ``Bubble period: 2003~-~2006'', ``US subprime loan crisis and the Great East Japan Earthquake: 2006~-~2012'', and ``BOJ monetary easing: 2013~-~present'', respectively.
The order parameters shows high spin ordering $M_{C,t} \approx M_{B,t} \approx 1$ during the bubble periods, and $M_{C,t} \approx M_{B,t} \approx -1$ during the crisis periods. 

In Fig. \ref{fig:OrderParameter}, we note that the phase synchronization was observed between 2002 and 2004, and between 2006 and 2008.
For these periods, the high spin ordering was observed in Figs. \ref{fig:mag_company} and \ref{fig:mag_bank}. 
The phase synchronization and high spin ordering are explained by the Kuramoto and Ising models, respectively, and are are interpreted as the collective motions in an economy.
This observation of the phase synchronization and high spin ordering in the same period supports the validity of the interacting agent models explained in Section \ref{sec:2}.

\subsection{Estimation of Exogenous Shock}

\begin{figure*}
  \includegraphics[width=0.8\textwidth]{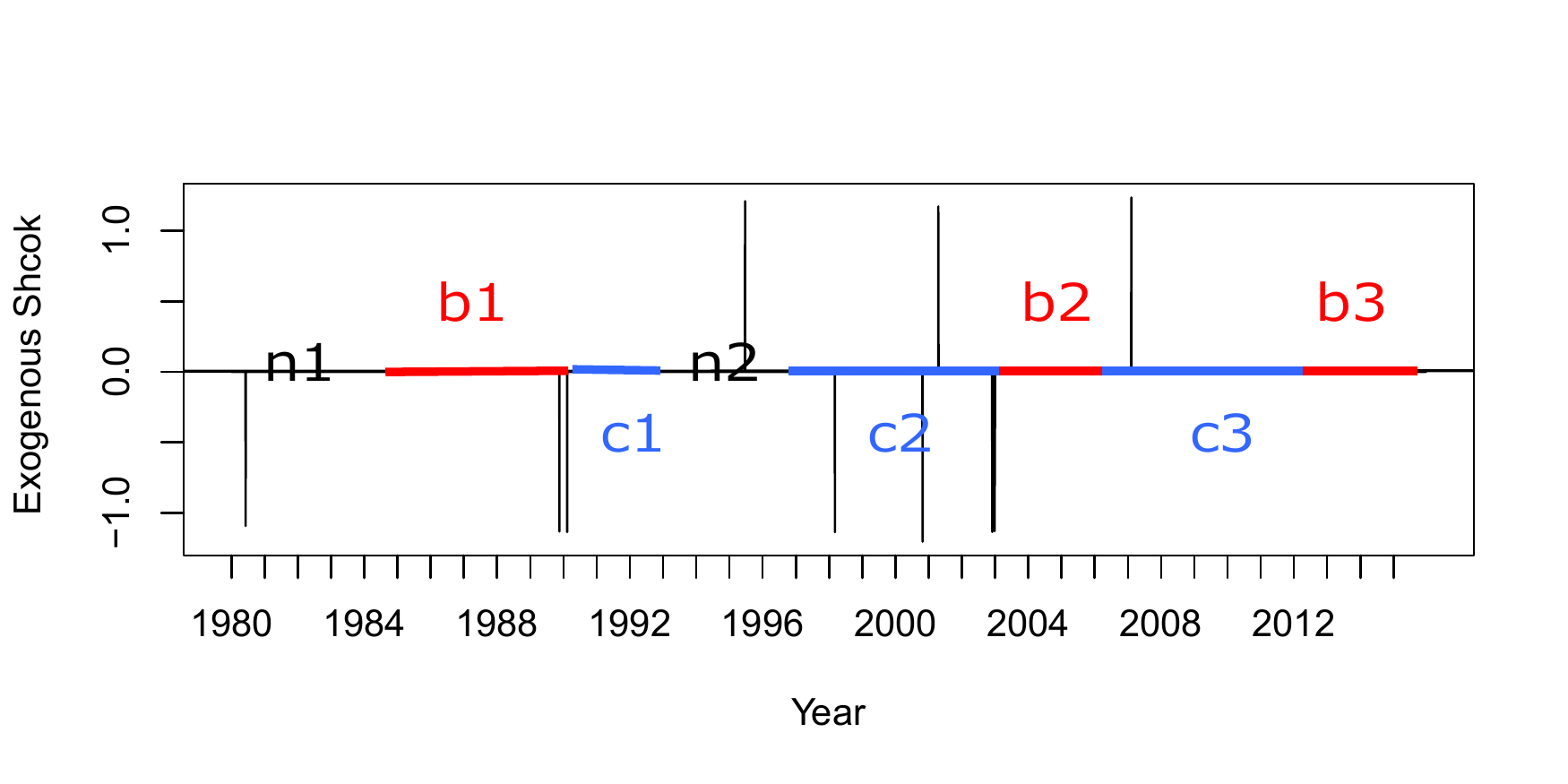}
\caption{Major mode of exogenous shock acting on the financial sector: The obtained exogenous shock acting on the financial sector indicates large negative shocks at the beginnings of c1 (1989) and c2 (1997), but no large negative shock during period c3 (2008). }
\label{fig:ext_bank}       % Give a unique label
\end{figure*}

Exogenous shocks were estimated using Eq. (\ref{eq:shock2}), and the major mode of an exogenous shock was extracted by eliminating shocks smaller than 90 \% of the maximal or the minimal shock. The major mode of exogenous shock acting on the financial sector is shown in Fig. \ref{fig:ext_bank}. The obtained exogenous shock acting on the financial sector indicates large negative shocks at the beginnings of c1 (1989) and c2 (1997) but no large negative shock during period c3 (2008). Therefore, the effect of the U.S. subprime loan crisis on the Japanese economy was introduced through shocks to the real economy (e.g., the sudden decrease of exports to the United States), not through direct shocks in the financial sector.

\begin{figure*}
  \includegraphics[width=0.8\textwidth]{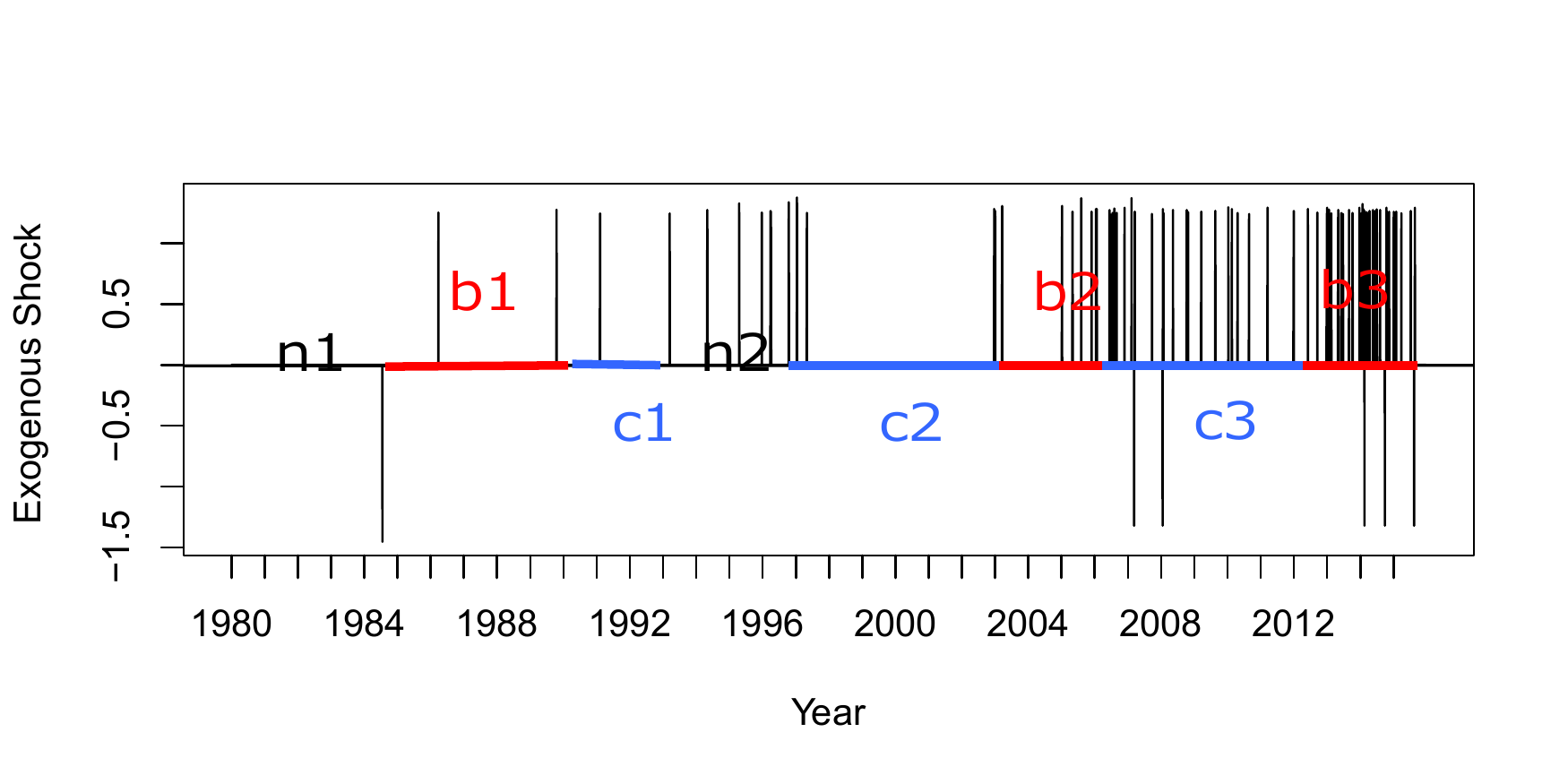}
\caption{Major mode of exogenous shock acting on the community, which consists of the construction, transportation equipment, and precision machinery sectors: The obtained exogenous shock acting on these sectors indicates no large negative shock at the beginnings of c1 (1989) and c2 (1997), but show a large negative shock during period c3 (2008).}
\label{fig:ext_comm4}       % Give a unique label
\end{figure*}

The major mode of an exogenous shock acting on the community, which consists of construction, transportation equipment, and precision machinery sectors, is shown in Fig. \ref{fig:ext_comm4}. For this community, no large negative shock was obtained at the beginnings of c1 (1989) and c2 (1997). 
In Fig. \ref{fig:mag_company}, we note that $M_{C,t} \approx −1$ is observed for this real economy community at the beginnings of c1 (1989) and c2 (1997). 
This observation is interpreted as the existence of channels of distress propagation from the financial sector to the real economy through the supply chain network in Japan. 
We observe a negative but insignificant exogenous shock on the real economy at the beginning of the U.S. subprime loan crisis (c3).

\subsection{Reconstruction of Interbank Network}

\begin{figure*}
  \includegraphics[width=0.5\textwidth]{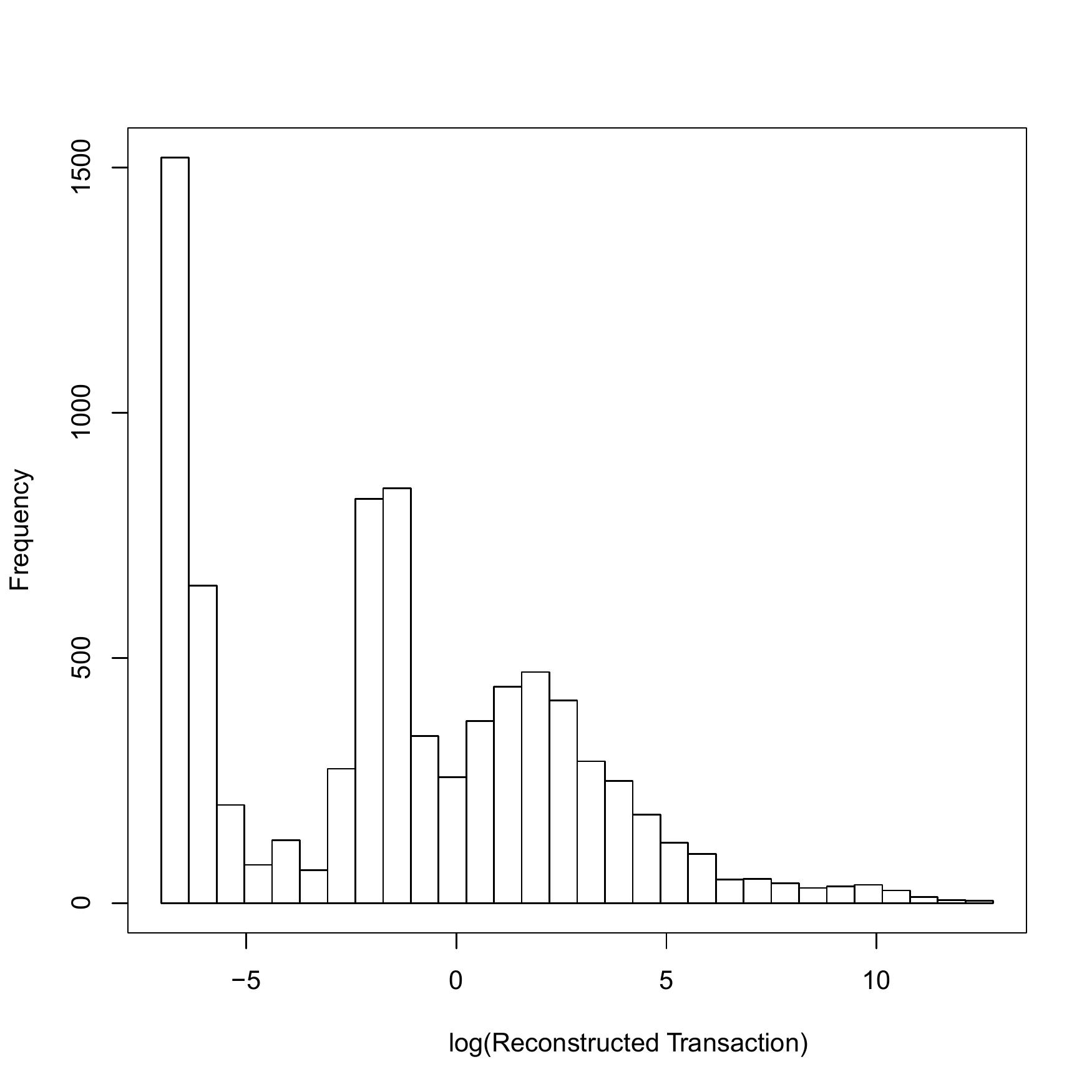}
  \includegraphics[width=0.5\textwidth]{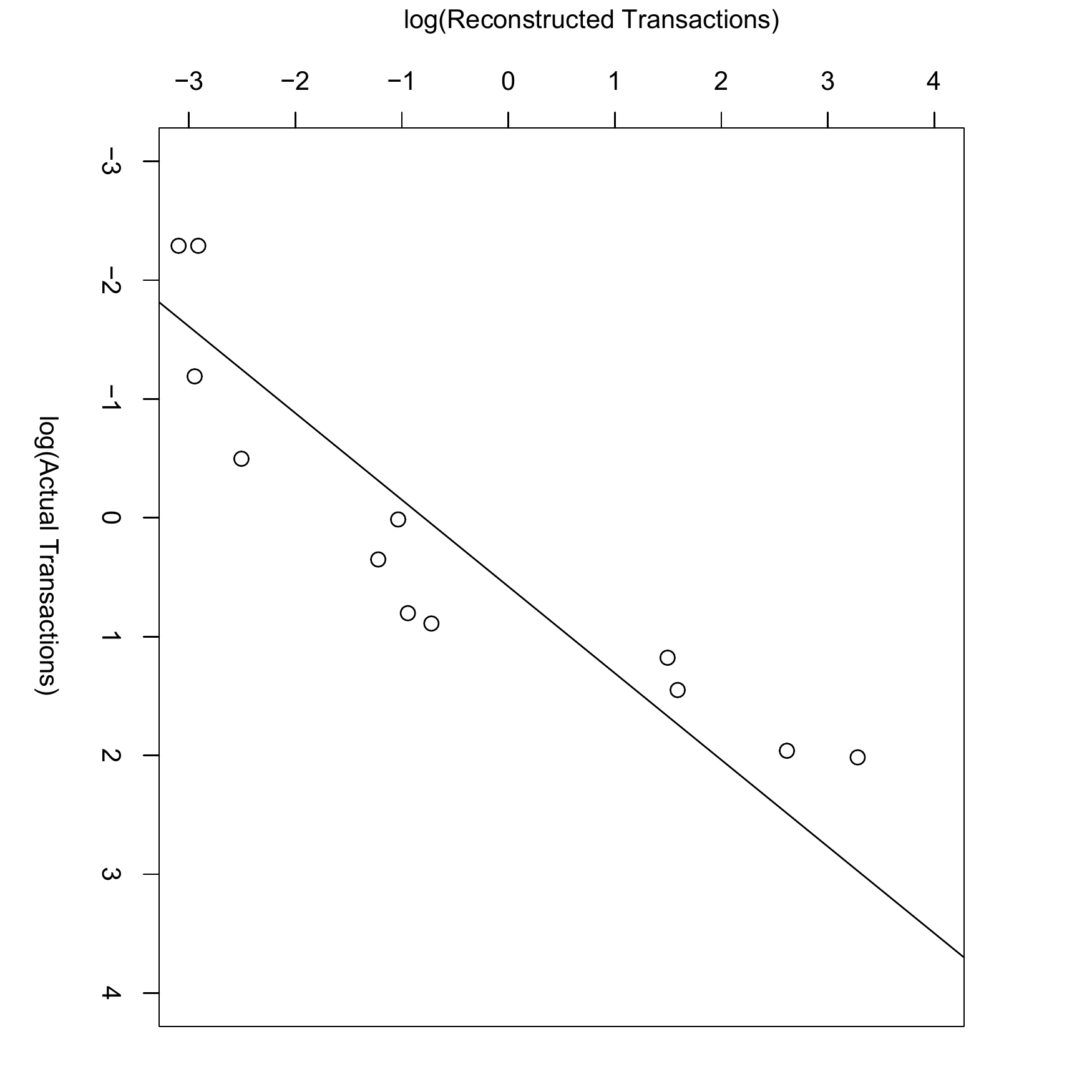}
\caption{Distribution of transactions $t_{ij}$ for the reconstructed interbank network in 2005 is shown in the left panel. A comparison of transactions between four categories of banks for the reconstructed interbank network and the actual values is shown in the right panel.}
\label{fig:EntropyModel_2005}       % Give a unique label
\end{figure*}

The interbank network in Japan was reconstructed using the ridge entropy maximization model in Eq. (\ref{ConvexFormulation}).
The number of banks in each category is 5, 59, 3, and 31 for major commercial bank, leading regional bank, trust bank, and second-tier regional bank, respectively. 
Call loan $s_{i}^{out}$ of bank $i$ and call money $s_{j}^{in}$ of bank $j$ are taken from each bank's balance sheet and are provided as constraints of the model.
In addition to the banks, a slack variable is incorporated into the model to balance the aggregated call loan and the aggregated call money. 
In the objective function in Eqs. (\ref{ConvexFormulation}), we assumed $\beta = 15$.

The distribution of transaction $t_{ij}$ for the reconstructed interbank network in 2005 is shown in the left panel of Fig. \ref{fig:EntropyModel_2005}. 
The leftmost peak in the distribution is regarded as zero and, thus, corresponds to spurious links.
Transactions $t_{ij}$ for the reconstructed interbank network were used to calculate the transaction between four bank categories and compared with the actual values taken from Table 4 in \cite{Imakubo2010}.
In the right panel of Fig. \ref{fig:EntropyModel_2005}, a comparison is shown of transactions among four categories of banks for the reconstructed interbank network and the actual values. 
This comparison confirms that the accuracy of the reconstruction model is acceptably good.

For the reconstructed interbank network, we obtain the following characteristics, which are consistent with the previously known stylized facts: 
the short path length, the small clustering coefficient, the disassortative property, and the core and peripheral structure.
Community analysis shows that the number of communities is two to three in a normal period and one during an economic crisis (2003, 2008~-~2013).
The major nodes in each community have been the major commercial banks.
Since 2013, the major commercial banks have lost the average PageRank, and the leading regional
banks have obtained both the average degree and the average PageRank.
This observed changing role of banks is considered to be the result of the quantitative and qualitative monetary easing policy started by the Bank of Japan in April 2013.

\section{Conclusions}
\label{sec:5}

Most national economies are linked by international trade. Consequently, economic globalization forms a massive and complex economic network with strong links, that is, interactions resulting from increasing trade. 
From the analogy of collective motions in natural phenomena, various interesting collective motions are expected to emerge from strong economic interactions in the global economy under trade liberalization. Among various economic collective motions, the economic crisis is our most intriguing problem.

We revealed in our previous studies that Kuramoto's coupled limit-cycle oscillator model and the Ising-like spin model on networks were invaluable tools for characterizing economic crises.
In this study, we developed a mathematical theory to describe an interacting agent model that derives these two models using appropriate approximations.
We have a clear understanding of the theoretical relationship between the Kuramoto model and the Ising-like spin model on networks.  
The model describes a system that has company and bank agents interacting with each other under exogenous shocks using coupled stochastic differential equations.
Our interacting agent model suggests the emergence of phase synchronization and spin ordering during an economic crisis.
We also developed a network reconstruction model based on entropy maximization considering the sparsity of network. 
Here, network reconstruction means estimating a network's adjacency matrix from a node's local information taken from the financial statement data.
This reconstruction model is needed because the central bank has yet to provide transaction data among banks to the public. 

We confirmed the emergence of phase synchronization and spin ordering during an economic crisis by analyzing various economic time series data.
In addition, the exogenous shocks acting on an industry community in a supply chain network and the financial sector were estimated. 
The major mode of exogenous shocks acting on a community, which consists of construction, transportation equipment, and precision machinery sectors was estimated. For this community, no large negative shock was obtained during the crises beginning in 1989 and 1997. 
However, negative spin ordering is observed for this real economy community during crises beginning in 1989 and 1997. 
Estimation of exogenous shocks acting on communities of the real economy in the supply chain network provided evidence of channels of distress propagation from the financial sector to the real economy through the supply chain network. 

Finally, we pointed out that, in our interacting agent model, interactions among banks were ignored because of the lack of transaction data in the interbank network. 
The interbank network was reconstructed using the developed model, and the reconstructed network and the actual data were compared. We successfully reproduce the interbank network and the known stylized facts.

%%%%%%%%%%%%%%%%%%%%%%%% referenc.tex %%%%%%%%%%%%%%%%%%%%%%%%%%%%%%
% sample references
% %
% Use this file as a template for your own input.
%
%%%%%%%%%%%%%%%%%%%%%%%% Springer-Verlag %%%%%%%%%%%%%%%%%%%%%%%%%%
%
% BibTeX users please use
% \bibliographystyle{}
% \bibliography{}
%

\end{document}